\documentclass[pra,twocolumn,aps]{revtex4} %
\usepackage{epsfig}
\makeatletter
\def\txtline#1{\noalign{\hbox{\strut\hskip\@totalleftmargin {#1}}}} %
\makeatother

\begin{document}
\bibliographystyle{prsty}

\title{Grover's Quantum Search Algorithm for an Arbitrary Initial Mixed State}
\author{Eli Biham and Dan Kenigsberg}
\affiliation{Computer Science Department, Technion, Haifa 32000, Israel.}

\begin{abstract}
The Grover quantum search algorithm is generalized to deal with an arbitrary
mixed initial state. The probability to 
measure a marked state as a function of time is calculated,
and found to depend strongly on the
specific initial state.
The form of the function, though, remains as it is in the case of initial
pure state.
We study the role of the von Neumann
entropy of the initial state, and show that
the entropy cannot be a measure for the
usefulness of the algorithm.
We give few examples and 
show that for some extremely mixed initial states (carrying high entropy),
the generalized Grover algorithm is considerably faster 
than any classical algorithm.\\
~\\
DOI: 10.1103/PhysRevA.66.062301 \hfill PACS number(s): 03.67.Lx
\end{abstract}

\maketitle 

\def\bra#1{\left<#1\right|}
\def\ket#1{\left|#1\right>}
\newcommand{\braOket}[3]{\langle #1|#2|#3\rangle}
\def\avg#1{\langle#1\rangle}

\section{Introduction}

Grover's search algorithm~\cite{Grover96,Grover97} provides an example of
the speed-up that would be offered by quantum computers, if and when they are
built. The problem solved by Grover's algorithm is 
finding a sought-after (``marked'')
element in an unsorted database of size \( N \). 
To solve this problem, a classical
computer would need $N/2$ database queries on average,
 and in the worst case it would need \( N-1 \) queries.
Using Grover's algorithm, a quantum computer can
find the marked state using only \( O(\sqrt{N}) \) quantum database queries.
The importance of Grover's result stems from the
fact that it proves the existence
of a gap (albeit a polynomial gap) between the power of quantum computers and
classical computers. Moreover, the algorithm may be used to speed up the
solution of many problems (such as NP-complete problems), 
for which no efficient classical algorithms is known.

Along this paper we assume without loss of generality that \( N=2^{n} \),
where \( n \) is an integer. The algorithm requires a register of \( n \)
qubits carrying the computation. When we say it is in a state \( \left| x\right\rangle  \),
we mean that its qubits are in states corresponding to the binary representation
of the number \( x \). Grover's original quantum search algorithm consists
of the following steps:

\begin{enumerate}
\item Initialize the register to \( H\left| 0\right\rangle  \). 
That is, reset all the
qubits to 0 and apply the Hadamard transform to each of them.
\item Repeat the following operation 
(named the \emph{Grover Iterate} $Q$) \( T=\frac{\pi \sqrt{N}}{4} \)
times:

\begin{enumerate}
\item Rotate the marked state $\ket k$ by a phase of
   $\pi$ radians $(I_{k}^{\pi})$. \label{alg:marked}
\item Apply the Hadamard transform to the register.
\item Rotate the $\ket{0}$ state by a phase of
   $\pi$ radians $(I_{0}^{\pi})$. \label{alg:pivot}
\item Apply the Hadamard transform again.
\end{enumerate} 

\item Measure the resulting state.
\end{enumerate}
Several generalizations extended the original Grover algorithm. Among these
is handling multiple marked states~\cite{BBHT96}, and the initialization of
the algorithm in any pure state~\cite{BB99}. Another generalization is the
replacement of the Hadamard transform by 
any other unitary operation~\cite{Grover98a,Ging00,BHMT00}.
In this way the algorithm may be used to 
speed-up many classical decision algorithms
and heuristics. Other generalizations use
arbitrary rotation angles~\cite{LLZT00},
replace the \( \left| 0\right\rangle  \) state from Step (c) with any other
state, or combine all of the above~\cite{BB01}. The rotation angles may be
tweaked in order to find a marked state with certainty~\cite{Hoyer00,LLS01}. 

The original Grover Iterate is $Q=-H I_0^{\pi} H I_k^{\pi}$.
It has been generalized
to $Q=-U I_s^{\beta} U^\dag I_M^{\gamma}$, 
where $U$ is an arbitrary unitary
operator, $s$ is an arbitrary state, 
$\beta$ and $\gamma$ are arbitrary angles, and $M$
includes any number of marked states.
We now observe that {\em any} unitary operation $Q$
has a unitary diagonalization.
Therefore, it can be represented as
 $Q=-U I_S^{\vec\beta} U^\dag I_M^{\gamma}$. This is a further generalization
of Grover's algorithm, where
the state $s$ is replaced by a set of states $S$, each of which 
may have a different rotation angle.
Thus, every iterative algorithm is a generalized Grover algorithm.

In this
paper we study the case where the generalized Grover Iterate of~\cite{BB01}
is applied to a quantum register that 
is initialized in an arbitrary mixed state. 
Our study extends and corrects a result from~\cite{Bose00}.

\section{Arbitrary Pure Initial State}
\label{sec:ArbPure}
If the abovementioned search algorithm is used as 
a procedure by another algorithm,
it might be necessary to avoid its first step. Even if the initialization is
performed, gate imperfection or external noise might cause the outcome to differ
from the exact $H\ket{0}$ state. Rather, it may
well be some general pure state \( \left| \psi _{0}\right\rangle  \), which
is a superposition of the marked state and the unmarked states. 
In addition, the Iterate itslef may be imperfect:
the Hadamard operation might be some other unitary operation $U$; 
the rotations of~(\ref{alg:marked}) and~(\ref{alg:pivot}) 
may be in angles $\beta$ and $\gamma$ (respectively), different of $\pi$;
and the rotated state of Step~(\ref{alg:pivot}) may be a non-zero $\ket s$.
Finally, the set of sought-after items, $M$, may include multiple items.

When the parameters of the problem are known, 
we may follow the results of Biham et al.~\cite{BB01}, 
and calculate the probability
to measure a marked state \( P_{\psi _{0}} \) as a function of the number of
Grover iterations $t$: 
\begin{equation}
P_{\psi _{0}}(t)=\left\langle P_{\psi _{0}}\right\rangle -\Delta P_{\psi _{0}}\cos (2\omega t+2\phi _{\psi _{0}}).
\end{equation}
$\avg{P_{\psi_0}}$ and $\Delta P_{\psi_0}$ denote the average over time and the
amplitude of $P_{\psi_0}$ respectively.
The subscripts \( \psi _{0} \) denote that the values depend on the initial
state.
However, \( \omega  \), which is defined by
\[\cos \omega = 
  \sum_{i\in M}\left|\braOket{i}{U}{s}\right|^2 \cos\frac{\beta+\gamma}{2}+
  \sum_{i\notin M}\left|\braOket{i}{U}{s}\right|^2 \cos\frac{\beta-\gamma}{2},
\]
is independent of the initial state.
In a large search problem with \( N\rightarrow \infty  \), 
and with the original Grover Iterate,
$\omega$ may be approximated by \( \omega =\frac{2}{\sqrt{N}} \). 
With the original initial state $H\ket0$
studied by Grover, we re-obtain 
\( \left\langle P_{H\left| 0\right\rangle }\right\rangle =
   \Delta P_{H\left| 0\right\rangle }=\frac{1}{2} \)
and \( \phi _{H\left| 0\right\rangle }\approx 0 \).

\section{Arbitrary Mixed Initial State}

A mixed state arises when one cannot describe the state of a quantum system
deterministically, no matter what basis one chooses. Such a condition appears
very often when a quantum system is entangled with its environment, while the
environment cannot be accessed or manipulated. The state of such a system may
be described by a completely-positive trace-1 hermitian density matrix, denoted
by \( \rho  \). An equivalent description is an ensemble \( {\cal E}=\{p_{\mu},\left| \psi _{\mu}\right\rangle \} \)
where \( \sum _{\mu}p_{\mu}=1 \) and \( \rho =\sum _{\mu}p_{\mu}\left| \psi _{\mu}\right\rangle \left\langle \psi _{\mu}\right|  \).
According to this description, the system is in the pure state \( \left| \psi _{\mu}\right\rangle  \)
with probability \( p_{\mu} \). 
When a unitary operation $V$ is applied to the mixed state
it transforms the state into \( \sum_{\mu}p_{\mu}V\ket{\psi_{\mu}}\bra{\psi_{\mu}}V^\dag  \).
The mixedness of the state does not change, and it may be thought as if $V$ 
transforms each of the components of $\cal E $ 
independently of the others.

Extending the argument of Section \ref{sec:ArbPure},
 the initial state of the quantum register
might not be pure, due to external noise, decoherence or previous manipulations.
Instead, the initial state may be some general mixed state \( \cal E \). Given
the description of $\cal E$ as an ensemble, all we can say is that the register
is in the pure state \( \left| \psi _{\mu}\right\rangle  \) with probability
\( p_{\mu} \) (for all $i$'s). 

\begin{figure}
\epsfig{file=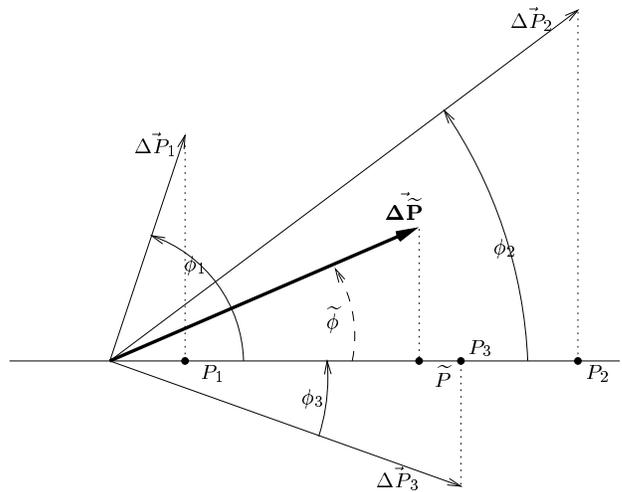}%
\caption{The change of probabilities as projections of rotating vectors.}
\label{fig:average}
\end{figure}

When the Grover algorithm is applied to a register 
which is in a pure state \( \ket{\psi_{\mu}} \),
the probability to measure the marked state is \( P_{\mu}(t) \). The probability
for the register to be in $\ket{\psi_{\mu}}$ is \( p_{\mu} \) 
(Considering the ensemble $\cal E$).
Thus, the total probability
to measure the marked state is the weighted average 
\begin{eqnarray} \label{weighted_average}
\widetilde{P}(t) & = & \sum _{\mu}p_{\mu}P_{\mu}(t) \nonumber\\ 
 & = & \sum _{\mu}p_{\mu}\left( {\vrule height2.5ex depth0pt width0pt} \avg{P_{\mu}}
        -\Delta P_{\mu}\cos \left( 2\omega t+2\phi _{\mu}\right) \right).%
\end{eqnarray}
The functions 
\[P_{\mu}(t)-\avg{P_{\mu}}=-\Delta P_{\mu}\cos \left( 2\omega t+2\phi _{\mu}\right)\] 
share a sinusoidal form, 
differing in amplitude
and phase, but not in frequency. 
They may be thought of as the projections of vectors rotating in frequency
$\omega$, as 
exemplified by Figure \ref{fig:average}.
Therefore, their weighted sum $\vec{\Delta\tilde P} $
(the Center of Mass of the vectors in the figure) is a
sinusoidal function with the same frequency:
\begin{equation}
\label{mixed_evolution}
\widetilde{P}(t)=\widetilde{\avg{P}}-\widetilde{\Delta P}\cos \left( 2\omega t+2\widetilde{\phi }\right) 
\end{equation}
while 
\[%
\widetilde{\avg{P}} =\sum _{\mu}p_{\mu}\avg{P_{\mu}},
\]%

\begin{equation}
\label{dP}
\widetilde{\Delta P}=\sqrt{\left( \sum _{\mu}p_{\mu}\Delta P_{\mu}\cos 2\phi _{\mu}\right) ^{2}+\left( \sum _{\mu}p_{\mu}\Delta P_{\mu}\sin 2\phi _{\mu}\right) ^{2}}
\end{equation}
and \begin{equation}
\label{phi}
\tan 2\widetilde{\phi }=\frac{\sum _{\mu}p_{\mu}\Delta P_{\mu}\sin \left( 2\phi _{\mu}\right) }{\sum _{\mu}p_{\mu}\Delta P_{\mu}\cos \left( 2\phi _{\mu}\right) }.
\end{equation}
 The probability to measure a marked state reaches its maximum value 
\[%
\widetilde{P}_{max}=\widetilde{\avg{P}}+\widetilde{\Delta P} 
\]%
after \( T=\frac{\pi -2\widetilde{\phi }}{2\omega } \) iterations. 

If the algorithm is repeated until success with 
$T$
iterations each time, the expected total time to measure a marked state is 
\[{\cal T}_Q=
  \frac{\pi -2\widetilde{\phi }}{2\omega \widetilde{P}_{max}}%
 \]
since the number of repetition until success is distributed geometrically with
parameter $\widetilde{P}_{max}$.
When the original Iterate is used, and a single item is sought after, this
reduces to 
${\cal T}_Q=\frac{\pi -2\widetilde{\phi }}{4 \widetilde{P}_{max}}\sqrt{N}.$ 
If this value is significantly smaller than the classical expected time 
\( {\cal T}_C=N/2 \), then
the quantum algorithm has an advantage.
Quantitatively, the expected number of oracle queries 
that the quantum algorithm requires 
is smaller by a factor of 
\begin{equation}
\label{condition}
\frac{{\cal T}_C}{{\cal T}_Q}=
  \frac{N \omega \widetilde{P}_{max}}{\pi -2\widetilde{\phi}}=
  \frac{2 \widetilde{P}_{max}\sqrt{N}}{\pi -2\widetilde{\phi }}.
\end{equation}
%

\section{Examples}
For clarity and simplicity, our examples use the original Grover Iterate
and single marked state $\ket k$, with
different initial mixed states.

\subsection{Pure Initial State}

When the arbitrary mixed state is chosen to be pure, the summations are degenerated
and the results of~\cite{BB99} are achieved. For example, if the initial state
is the original \( {\cal E}=\{p=1,H\left| 0\right\rangle \} \), the original
Grover case is found. If \( {\cal E}=\{p=1,\ket k \} \),
then \( \widetilde{\avg{P}}=\widetilde{\Delta P}=\frac{1}{2} \) and \( \widetilde{\phi }=\frac{\pi }{2} \).
An interesting known property of the Grover algorithm is that for all states
orthogonal to both \( \ket k \) and \( H\left| 0\right\rangle  \),
\( \widetilde{\avg{P}}=\widetilde{\Delta P}=0 \).

\subsection{Pseudo-Pure Initial State}

Ensembles where a state $\ket{\psi}$ appears with probability 
$\epsilon+\frac{1-\epsilon }{N} $ and any
state orthogonal to it appears with 
equal probabilities of $\frac{1-\epsilon }{N}$
are called pseudo-pure mixed states. They are written more conveniently as 
\( \rho _{\epsilon \mbox{-}pure}=(1-\epsilon)\frac IN+\epsilon \ket{\psi}\bra{\psi} \).
Notice that \( 0\leq \epsilon \leq 1 \) is a measure
of the purity of \( \rho  \): when \( \epsilon =0 \) it is totally mixed,
and when \( \epsilon =1 \) it is totally pure. It is easy to see that in the
limit of large $N$, \( \widetilde{\avg{P}} =\epsilon \left\langle P_{\psi }\right\rangle  \),
\( \widetilde{\Delta P}=\epsilon \Delta P_{\psi } \) and \( \widetilde{\phi }=\phi _{\psi } \).
For example, for 
\[ \rho _{\frac{1}{\log N}\mbox {-}pure}=(1-\frac{1}{\log N})\frac IN+\frac{1}{\log
N}H\left| 0\right\rangle \left\langle 0\right| H, \]
we obtain \( \widetilde{\avg{P}} =\widetilde{\Delta P}=\frac{1}{2\log N} \)
and \( \widetilde{\phi }=0 \). 
Notice that although \( \rho  \) is extremely
mixed, the quantum advantage is of factor 
\( \frac{2\sqrt{N}}{\pi \log N} \).

\subsection{Initial State Where $m$ of the Qubits Are Mixed}

Let us study the case where the register is initialized to \( \rho _{m\mbox {-}mix}=2^{-m}\sum ^{2^{m}-1}_{i=0}H\left| i\right\rangle \left\langle i\right| H \).
This state may occur if the \( m \) least significant qubits of the register
are totally mixed before the first Hadamard transform is applied. Since all
\( H\left| i\right\rangle  \) are orthogonal to \( H\left| 0\right\rangle  \)
(except for \( H\left| 0\right\rangle  \) itself) and they are almost orthogonal
to \( \ket k \) (since \( \left| \left\langle k\right| H\left| i\right\rangle \right| ^{2}=\frac{1}{N} \)),
the evolution of \( \rho _{m \mbox{-}mix} \) is governed by \( \{p=2^{-m},H\left| 0\right\rangle \} \)
and we obtain \( \widetilde{\avg{P}}=\widetilde{\Delta P}=\frac{1}{2^{m+1}} \)
and \( \widetilde{\phi }=0
 \). Large $m$ would render the
algorithm useless.

\section{Algorithm Usefulness and Entropy}

The von Neumann entropy of a mixed state $\rho$ is defined as 
$S(\rho)=-\mbox{tr}\rho \log \rho$.
Bose et al.~\cite{Bose00} presented a new model for quantum computation
and laid out a new proof for the optimality of the Grover algorithm. However,
one of their results was the following: if the Grover algorithm is initiated
with a mixed state $\rho$, such that \( S(\rho )\geq \frac{1}{2}\log N \),
the algorithm would have no advantage comparing to the classical case. This
is in disagreement with our findings~\cite{Bose00err}. 
A counter-example to their claim is \( \rho _{\frac{1}{\log N}\mbox {-}pure} \)
as defined above. The entropy of pseudo-pure state is
\begin{eqnarray}
S(\rho _{\epsilon \mbox {-}pure}) & = & 
	S\left( \frac{1-\epsilon }{N}I_{N}+
                \epsilon \ket{0}\bra{0}\right) \nonumber\\
 & = & -\sum _{1}^{N-1}\frac{1-\epsilon }{N}\log \frac{1-\epsilon }N \nonumber\\
 &   & ~~~~~~~-\frac{1+(N-1)\epsilon}N\log\frac{1+(N-1)\epsilon }N\nonumber\\
 & = & -(N-1)\frac{1-\epsilon }{N}\log \frac{1-\epsilon }{N} \nonumber\\
 &   & ~~~~~~~-\frac{1+(N-1)\epsilon}N\log \frac{1+(N-1)\epsilon}N,\nonumber\\
\txtline{and for large $N$, where $N/(N-1)\approx1$,}
 &\approx& -(1-\epsilon )\log \frac{1-\epsilon }N
      -\left(\frac1N+\epsilon\right)\log\left(\frac1N+\epsilon\right)\nonumber\\
 & = & (1-\epsilon)\log N -(1-\epsilon )\log (1-\epsilon )\nonumber\\
 &   & ~~~~~~~ -\left(\frac1N+\epsilon\right)\log\left(\frac1N+\epsilon\right)\nonumber\\
 &=& (1-\epsilon)\log N-\ell, \label{eq:PsPureEnt}
\end{eqnarray}
where $\ell = 
   (1-\epsilon)\log(1-\epsilon)+\left(\frac1N+
   \epsilon\right)\log\left(\frac1N+\epsilon\right) 
 \in (-1,0.8)$ for any $0\leq\epsilon\leq1$ and any $N\geq 2$.
For \( \epsilon =\frac{1}{\log N} \) we obtain 
\( S(\rho _{\frac{1}{\log N}\mbox {-}pure})=
   \left( 1-\frac{1}{\log N}\right) \log N+O(1)=\log N+O(1) \).
This entropy is almost maximal. However, as noted above, the Grover algorithm
outperforms any classical algorithm, 
even when it is initialized with this state.

Entropy is not a good measure for the usefulness of Grover's algorithm. 
For practically every value of entropy,%
there exist states that are good initializers and
states that are not. For example, 
\( S(\rho _{(n-1)\mbox {-}mix})=\log N-1=
   S(\rho _{\frac{1}{\log N}\mbox {-}pure}) \),
but when initialized in \( \rho _{(n-1)\mbox {-}mix} \), the Grover algorithm
is as bad as guessing the marked state. Another example may be given using the
pure states \( H\left| 0\right\rangle \left\langle 0\right| H \) and \( H\left| 1\right\rangle \left\langle 1\right| H \).
With the first, Grover arrives to the marked state with quadratic speed-up,
while the second state is practically unchanged by the algorithm.

\section{Acknowledgments}

We thank Tal Mor for valuable discussions and suggestions that
made the compiling of this paper possible.
The work was partially supported by the European
Commission through the IST Programme under contract IST-1999-11234.
The first author was partially supported by the
fund for the promotion of research at the Technion
and the Israel MoD Research and Technology Unit.

\end{document}